\documentclass[creativecommons]{eptcs}
\providecommand{\event}{E. Formenti (Ed.): AUTOMATA \& JAC 2012} 
\usepackage{makeidx,amsfonts,graphicx,amsmath}
\def\zz{{\mathbb Z}}

\def\nn{{\mathbb N}}
\def\ff{{\mathbb F}}
\newcommand{\ul}{\underline}
\newcommand{\bt}{\begin{tabular}}
\newcommand{\et}{\end{tabular}}
\newcommand{\ba}{\begin{array}}
\newcommand{\ea}{\end{array}}
\newcommand{\bq}{\begin{eqnarray}}
\newcommand{\eq}{\end{eqnarray}}
\newcommand{\bp}{\begin{proof}}
\newcommand{\ep}{\end{proof}}
\newcommand{\sgn}{\operatorname{sgn}}
\providecommand{\urlalt}[2]{\href{#1}{#2}} 
\providecommand{\doi}[1]{doi:\urlalt{http://dx.doi.org/#1}{#1}}

\begin{document}

\title{Computing by Temporal Order:\\
Asynchronous Cellular Automata}

\author{Michael Vielhaber
\thanks{Partially funded by HS Bremerhaven, Germany through a
  sabbatical leave.}  
\institute{
Universidad Austral de Chile,
  Instituto de Matem\'aticas, Casilla 567, Valdivia, Chile}
\institute{
Hochschule Bremerhaven, FB2, An der Karlstadt 8,
D--27568 Bremerhaven, Germany}
\email{vielhaber@gmail.com}
}

\def\titlerunning{Computing by Temporal Order}
\def\authorrunning{M.~Vielhaber}

\maketitle

\begin{abstract}
Our concern is the behaviour of the elementary cellular automata with
state set $\{0,1\}$ over the cell set $\zz/n\zz$
(one-dimension\-al finite wrap-around case), under {\it
  all} possible temporal rules (asynchronicity). 

Over the torus $\zz/n\zz$ $(n\leq 10)$,we will see that the ECA with Wolfram
update rule 57 maps any $v\in \ff_2^n$ to any   
$w\in \ff_2^n$, varying the temporal rule.
 
We furthermore show that all even (element of the alternating group)
bijective functions on the set  
$\ff_2^n\cong\{0,\dots,2^n-1\}$, 
can be computed by ECA-57, by iterating it a sufficient number of times
with varying temporal rules, at least for $n\leq 10$.  
We characterize the non-bijective functions computable by asynchronous rules.

The thread of all this is a novel paradigm:

The algorithm is neither hard-wired (in the ECA), nor in the program
or data (initial configuration), but in the temporal order of updating
cells, and temporal order is pattern-universal. 

{\bf Keywords:}
Cellular automata, asynchronous, update rule,
universality.
\end{abstract}

\section{Introduction and Notation, Asynchronicity}

We consider elementary cellular automata, {\it i.e.} with state set
$S=\ff_2=\{0,1\}$ and update neighborhood  $(c_{i-1},c_i,c_{i+1})$ for cell
$c_i$.

The cell index (site) $i$ will come  from $\zz/n\zz$ for some $n\geq
3$, {\it i.e.} we consider the  finite one-dimensional torus, indices
wrap around.
In Section 2, we consider patterns ``How universal can a mapping
 on $\ff_2^n$ become?'',  and Section~3 covers
functions $\ff_2^n\ni v\to w\in \ff_2^n$.

The 256 ECA's group into 88 classes under the symmetries 0/1 
and left/right neighbor, see Appendix~A.
It is sufficient to consider one member per class.

The Wolfram rule  ECA = $\sum_{k=0}^7 2^k\cdot p_k\in\{0,\dots,255\}$
defines the behavior.
A cell with neighborhood $(c_{i-1},c_i,c_{i+1})$ $\in \ff_2^3$,
summing up to $k:=4c_{i-1}+2c_{i}+c_{i+1}\in\{0,\dots,7\}$
is replaced by $c_i^+ := p_k$.

{\bf Example~1}\ \
The behaviour of the ECA with Wolfram rule $57_{10}={\bf 00111001}_2$
is given in Table~1.  
We have that 
$0c_i1\mapsto c_i$, all other cases
$0c_i0, 1c_i0,1c_i1\mapsto \overline c_i$. 

\begin{table}[h]
\centering
$111\mapsto {\bf 0}$,\quad \quad \  $011\mapsto {\bf 1}$,\\            
$110\mapsto {\bf 0}$,\quad \quad \  $010\mapsto {\bf 0}$,\\           
$101\mapsto {\bf 1}$,\quad \quad \  $001\mapsto {\bf 0}$,\\            
$100\mapsto {\bf 1}$,\quad \quad \  $000\mapsto {\bf 1}$.\\
\caption{ECA-57}
\end{table}

\subsection{State-of-the-Art}

The study of asynchronous cellular automata started with Ingerson and
Buvel's  1984 paper \cite{IB}.  

Lee {\it et al.} \cite{Lee} give an asynchronous CA on the
two-dimensional grid $\zz\times \zz$, which is Turing-universal. 

Fat\`es {\it et al.} \cite{Fates} consider  ECA's with
quiescent states ($000\mapsto 0, 111\mapsto 1$, {\it i.e.} with even
Wolfram rule $\geq 128$). They consider fully randomized ECA's. 

A good overview is given in the thesis \cite{Sharkar} by Sharkar.

Nevertheless, all these articles treat asynchronous CAs with
randomized clocks.

Our concern is instead the (fully deterministic) behavior of a suitable ECA,
with any fixed initial configuration, under {\it all} possible temporal
sequences. 
There seems to be no work on the combined effect of all deterministic
temporal rules,  synchronous and asynchronous, so far.

{\it Definition~1.} Temporal Rules --- Asynchronicity Rules

Let the set ${AS}_n$ of asynchronicity rules over $\zz/n\zz$ consist
of all words of length $n$ over the alphabet $\{<,\equiv,>\}$ such
that both $<$ and $>$ occur at least once. We also include the word 
``$\equiv\cdots\equiv$'', the synchronous case, and have
${AS}_n = 
(\{<,\equiv,>\}^n\backslash
(\{<,\equiv\}^n\cup
\{\equiv,>\}^n))
\cup \{\equiv^n\}$ 
with $|AS_n|=3^n-2^{n+1}+2$.

Given a rule $\mbox{\sc as}=\mbox{\sc as}_0\cdots\mbox{\sc as}_{n-1}$,
$\mbox{\sc as}_i =$ ``$<$'', ``$\equiv$'', and ``$>$'', resp., 
defines that cell $c_i$ updates
after,  simultaneously with, resp.~before $c_{i+1}, \forall 0\leq i\leq n-2$. 
$\mbox{\sc as}_{n-1}$ refers to cell $c_{n-1}$ with respect to $c_0$.

For any partition $(S_1,\dots S_m)$ of the cell sites, {\it i.e.} 
$\dot\cup_{k=1}^m S_k=\{0,\dots,n-1\},$
let its temporal rule be
$\mbox{\sc as}_i = \ <, \equiv, >\ $, resp.,  if $i\in S_{\iota(i)}$,  
$i+1\in S_{\iota(i+1)}$, and 
$\iota(i) $ is $>, =, < $, resp., than $\iota(i+1)$ 
(we say that site $i$ is ``bigger''
if it comes before $i+1$, hence dominates it).

With the exception of $\equiv^n$ (synchronous case), both $<$ and $>$ must
occur at least once, since otherwise, by wrapping-around, each cell
would update only after itself and the temporal rule would thus not
be well-defined, {\it e.g.} $\equiv<\equiv$ leads to $c_1$ with $c_2$ after
$c_3$ with $c_1$, so $c_1$ after, and thus before, itself.

{\bf Example~2}
Let $n=4$, and {\sc as} = ``$<\equiv>>$'':
Cell 0 updates after cell 1, 1 with 2, 2 before 3, and 3 before 0.
Hence the temporal order is $(1,2|3|0)$, first 1 and 2 simultaneously,
then 3, finally cell 0, {\it i.e.} $S_1=\{1,2\},S_2=\{3\},S_3=\{0\}$. 
Analogously,  ``$<><>$'' leads to $(1,3|0,2)$, and
``$>\equiv><$'' leads to $(0|1,2|3)$.

One might be inclined to partition the $n$ cells into
sets $S_1,\dots,S_m\subset \zz/n\zz$, and update those in $S_1$ first,
then cells from $S_2$ and so forth. 
This, however, is too fine-grained:

{\bf Theorem~1}

{\it
Consider two partitions  $(S_1,\dots, S_m)$ and  $(S'_1,\dots, S'_{m'})$
of the cell set $\zz/n\zz$ and define functions $a,b,c,a',b',c'$ such that
$$\forall i\in \{0,\dots,n-1\}:
i-1\in S_{a(i)}, i\in S_{b(i)},i+1\in S_{c(i)},\dots\quad\quad\quad\quad$$
$$\quad\quad\quad\quad\dots i-1\in S'_{a'(i)}, i\in S'_{b'(i)},i+1\in S'_{c'(i)}.$$
Then, if $\sgn(a(i)-b(i)) = \sgn(a'(i)-b'(i))$
and  $\sgn(b(i)-c(i)) = \sgn(b'(i)-c'(i)), \forall i$, {\it i.e.} the
relative update
order of cells $i-1,i,i+1$ is the same for $S$ and $S'$,
then updating according to $S$ or according to $S'$ leads to the same
result, and this is described by the following asynchronicity rule (Table~2).

\begin{table}[h]
\centering
$\ba{cc|cc|l}
\sgn(a-b)&\sgn(b-c)&\mbox{\sc as}_{i-1}&\mbox{\sc as}_{i}\\
\cline{1-5}
-1&-1&>&>&\mbox{\rm\ a\ before\ b\ before\ c}\\
-1&\ 0&>&\equiv &\mbox{\rm\ a\ before\ b\ with\ c}\\
-1&+1&>&<&\mbox{\rm\ a\ and\ c\ before\ b}\\
\ 0&-1&\equiv &>&\mbox{\rm\ a\ with\ b\ before\ c}\\
\ 0&\ 0&\equiv &\equiv &\mbox{\rm\ a\ with\ b\ with\ c}\\
\ 0&+1&\equiv &<&\mbox{\rm\ c\ before\ a\ with\ b}\\
+1&-1&<&>&\mbox{\rm\ a\ before\ b\ and\ c}\\
+1&\ 0&<&\equiv &\mbox{\rm\ b\ with\ c\ before\ a}\\
+1&+1&<&<&\mbox{\rm\ c\ before\ b\ before\ a}\\
\ea$
\caption{Local asynchronicity}
\end{table}
}

{\it Proof.}
By construction.
Since the relative temporal order of cell $c_i$ with respect to $c_{i-1}$ and 
$c_{i+1}$ is the same for $(S_k)$ and $(S'_k)$ by  
$\sgn(a(i)-b(i)) = \sgn(a'(i)-b'(i))$
and  $\sgn(b(i)-c(i)) = \sgn(b'(i)-c'(i))$,
both partitions lead to the same overall behaviour,
which is described by  $\mbox{\sc as}$.\hfill$\Box$

The construction by the theorem shows that the $\mbox{\sc as} \in
  AS_n$ are sufficient to distinguish the behaviour. On the other
hand, all these {\sc as} are necessary and can lead to different
behaviour (at least for some ECA's), since any $\mbox{\sc as}_i\neq \mbox{\sc
  as'}_i$ will lead to a different order of updating cells $c_i$ and $c_{i+1}$.

{\bf Example~3}\ \ 
For $n=6$, ``$<><><>$'' requires the odd cells $1,3,5$ to update before the even
ones $0,2,4$. There are 13 partitions of three elements, {\it e.g.}
$(1,3,5), (1|3,5),$ $(1,5|3)$, and $(5|1|3)$, and thus $13^2=169$ 
partitions $(S_k)$ for this {\sc as}.

{\it Definition~2.} \ \ 
By $\mbox{\rm ECA}_{\mbox{\sc as}}(v) = w$, we mean that the elementary
CA with rule {\rm ECA} maps $v\in\{0,1\}^n$ to $w\in\{0,1\}^n$ via
the temporal sequence {\sc as}.

{\bf Example~4}\ \ 
ECA-57$_{<><>}(1000) = 1110$, in two steps: $1\ul 00\ul 0\mapsto
\ul 11\ul 00\mapsto 1110$, where underlined cells are active in the
next step.

\section{The Finite Torus $\zz/n\zz$:
Patterns}

In this section, we work on the torus $\zz/n\zz$, and 
consider all ECA's for all initial configurations. 
We apply a fixed temporal rule $\mbox{\sc as}\in {AS}_n$ repeatedly,
$\tau$ times, and ask, whether these 5 pattern universality properties hold:

$\left.\ba{clll}
(o)&\exists v\in \ff_2^{n},&\forall w\in \ff_2^{n}, &\exists \tau\in\nn, \dots
\\
(i)&\forall v\in \ff_2^{n}, &\forall w\in \ff_2^{n}, &\exists \tau\in\nn,\dots
\\
(ii)&\forall v\in \ff_2^{n}, &\exists \tau\in\nn,&\forall w\in \ff_2^{n}, \dots
\\
(iii)&\exists \tau\in\nn,&\forall v\in \ff_2^{n},&\forall w\in \ff_2^{n},\dots
\\
(iv)&\exists \tau_0\in\nn,&\forall \tau\geq \tau_0,&\forall v,w\in
\ff_2^{n},\dots  
\ea\right\}\ \exists \mbox{\sc as}\in{AS}_{n}:
\mbox{\rm ECA}^\tau_{\mbox{\sc as}}(v) = w$.

All results are experimental {\it i.e.}  derived from exhaustive computer
simulations for the stated lengths.

We start with

$(o)$ $\exists v\in \ff_2^{n},\forall w\in \ff_2^{n}, \exists
\tau\in\nn, \exists \mbox{\sc as}\in{AS}_{n}: 
\mbox{\rm ECA}^\tau_{\mbox{\sc as}}(v) = w$.
That is from some $v$ we eventually reach any $w$.
We give the largest
number of $w$'s reached for some $v$, for $n=4, 8$, and $12$. 
To satisfy~$(o)$, these must be $(16, 256, 4096)$.

\noindent The 3 ECA families 0 (1,1,1),  200 (1,1,1), and  204
(1,1,1) are resilient to asynchronicity. 
They have a constant result, for all  $n$. 

\noindent ECA-51 (2,2,2) varies between at most two results.

The next 49 ECA families are ordered by increasing image size for $n=12$:

\nopagebreak

\noindent{
\bt{rlrlrlrlr}
140      &(2,6,16),& 160 &(12,130,1182), & 164 &(13,197,2930), & 108 &(16,256,4052),& \\     
136      &(2,9,27),&   2 &(11,211,1477), &  24 &(15,211,2961), &  56 &(16,256,4066),& \\
128     &(2,16,49),&  72 &(11,131,1499), &  34 &(13,209,2998), &  74 &(15,255,4071),& \\
132     &(4,18,81),&  76 &(11,131,1499), & 130 &(14,211,3160), &  73 &(16,256,4084),& \\
 32    &(7,31,127),& 172 &(11,137,1506), &  94 &(16,216,3448), &  33 &(16,256,4092),& \\
  8    &(5,45,320),& 168 &(12,147,1601), & 152 &(14,237,3561), &  10 &(13,253,4093),& \\
  4    &(7,47,322),&  13 &(16,168,1792), & 138 &(13,238,3751), & 134 &(15,255,4093),& \\
 12    &(7,47,322),& 232 &(12,156,1830), & 104 &(14,232,3824), &  42 &(15,255,4093),& \\
 28   &(11,91,641),&  77 &(12,156,1830), & 162 &(16,250,3970), &  35 &(16,256,4094),& \\
 29   &(12,92,642),& 142 &(12,140,1848), & 170 &(16,256,3976), &  43 &(16,256,4094),& \\
 44  &(12,100,870),&  78 &(15,167,1851), &  15 &(16,256,3976), &                   & \\
156   &(4,64,1024),&  36 &(14,162,1943), & 150 &(12,240,4032), &                   & \\
 40 &(11,119,1052),&   5 &(16,216,2542), &   1 &(16,256,4051), &                   & \\
\et
}
 
The 4 ECA families 6, 14, 18 [for $n\geq 7$], 50 [for $n\geq 4$],
miss exactly one pattern, leading to $2^n-1$ in general.

Finally, the 31 ECA families\\ 
3, 7, 9, 11, 19, 22, 23,
25, 26, 27, 30, 37, 38, 41, 45, 46, 
54, 57, 58, 60, 62, 90,\\ 
105 [$n\neq 0 \mod 4]$,
106, 110, 122, 126, 146, 154, 178 [$n\neq 3]$, 184 [$n\neq 3]$,\\
satisfy property $(o)$ (for $3\leq n \leq 12$).

$(i)$ $\forall v,w\in \ff_2^{n}, 
\exists \mbox{\sc as}\in{AS}_{n}, \exists \tau\in\nn: 
\mbox{\rm ECA}^\tau_{\mbox{\sc as}}(v) = w$.
\nopagebreak
From the 31 families satisfying $(o)$, most fall short for some $v$.
We give the smallest number of $w$ reachable from some $v$, for $n=4,
8,$ and 12, this should be (16,256,4096) to satisfy $(i)$.
 
Eighteen ECA families are insensitive (or resilient) to
asynchronicity for at least some 
$v$,  the same $w$ resulting for all \mbox{\sc as}.
Hence, $(1,1,1)$ patterns are reached:\\ 
22, 26, 30, 38,  46, 
54, 58, 60, 62, 
90, 106, 110, 
122, 126, 146, 154, 
178, 184.

\noindent ECA family 7 reaches $2^n-1$ for $n\neq 0\mod3$ and only 1 pattern
for $n\equiv 0\mod 3$.\\
ECA family 45 has $2^n-1$ patterns for odd $n$, 1 for even $n$.
 
\noindent Six ECA families get near the full $2^n$ for all $w$:\ 
3 (15,233,3411),  9(12,243,3963), 11 (15,233,3515),
25 (16,251,4031),  27 (16,253,4052), 43 (12,236,3554).

The following 6 ECA families satisfy 
$(i)$ at least for certain $[n]$ ($3\leq n\leq 12$ considered):

19 [3-12], 23 [3,5,7,9,11], 
37 [4-5,7-8,10-11], 
41[3,5,7-12],  57[3-12], 105 [3,5-7,9-11] all generate
$2^n$ patterns  for these $[n]$.

$(ii)-(iv)$ From now on, we will consider the 6 ECA families
satisfying $(i)$: 
19, 23 $(n\not\equiv 0 \mod 2)$, 
37 $(n\not\equiv 0 \mod 3)$, 41, 57, and
105 $(n\not\equiv 0 \mod 4)$.

$(ii)$ $\forall v\in \ff_2^{n}, \exists \tau\in\nn,\forall w\in
\ff_2^{n},\exists \mbox{\sc as}\in{AS}_{n}: 
\mbox{\rm ECA}^\tau_{\mbox{\sc as}}(v) = w$;
{\it i.e.} for fixed $v$, all $w$ are  reached at the same time.

We considered $\tau$ up to 20000, and obtain:

\noindent ECA-23: No $v$  has any $\tau\leq 20000$ to satisfy $(ii)$.

\noindent ECA-19,-37,-41: For some $v$, there is no $\tau \leq 20000$ to
satisfy $(ii)$. 

\noindent ECA-57 satisfies $(ii)$, for $n\geq 5$ and all $v$.
The largest $\tau$ required is $28$ for $n=5$; $14$ for $n=6$;
$10$ for $7\leq n\leq 13$; and 9 for $n=14$ and 15. 

\noindent ECA-105 satisfies $(ii)$ for odd $n\geq 7$ and all $v$.
The largest $\tau$ required is  $30$ for $n=7$; $16$ for $n=9,11,13$; and
$8$ for $n=15$. 

In general, the time $\tau$ decreases with $n$, since
the num\-ber of patterns, $2^n$, increases slower than the number of
asynchronicities,  $3^n$--$2^{n+1}$+$1$, and thus for larger $n$, 
${AS}_n$ is more likely to satisfy $(ii)$ early on.

$(iii)$ $\exists \tau\in\nn,\forall v,w\in \ff_2^{n},\exists \mbox{\sc
  as}\in{AS}_{n}: 
\mbox{\rm ECA}^\tau_{\mbox{\sc as}}(v) = w$. 
All transductions $v\mapsto w$ are  done in the same time. 

From the result of $(ii)$, we can infer that at most 
ECA-57 and ECA-105 can satisfy $(iii)$.

\noindent ECA-57 has a joint $\tau_n$ at which all transductions $v\mapsto w$ are
satisfied simultaneously in these cases:
$\tau_5=445,\tau_7=70,\tau_8=242,\tau_9=35,\tau_{10}=\cdots=\tau_{14}=13,$ 
$\tau_{15}=10$.

\noindent For ECA-105, we have  $\tau_7=570,\tau_9=14,\tau_{11}=\tau_{13}=6,$ 
and $\tau_{15}=8$.

$(iv)$ $\exists \tau_0\in\nn,\forall \tau\geq \tau_0,\forall v,w\in
\ff_2^{n},\exists \mbox{\sc as}\in{AS}_{n}: 
\mbox{\rm ECA}^\tau_{\mbox{\sc as}}(v) = w$.
Eventually all
transductions $v\mapsto w$ can be done at all times.

{\bf Theorem~2}

{\it
There is no $\tau_0\in\nn,\forall \tau\geq \tau_0,\forall v,w\in
\ff_2^{n},\exists \mbox{\sc as}\in{AS}_{n}: 
\mbox{\rm ECA}^\tau_{\mbox{\sc as}}(v) = w$, {\it i.e.} $(iv)$ can not
be satisfied. 
}

{\it Proof.}
Consider the case $w=v$.

For each rule {\sc as}, applying {\sc as} repeatedly, starting at $v$,
we will either return to $v$ at some time, which is the period length
 $per(\mbox{\sc as})$, the length of the cycle of {\sc as} containing
$v$, or else $v$ is in a preperiod and will never be reached again.
Therefore,
either $v$ is in the preperiod and thus will
not reappear, or else there is no preperiod, and $v=w$ appears exactly
after $k\cdot per(\mbox{\sc as}),\forall k$, applications of  {\sc
  as}.

Let now $\operatorname{PER} = lcm_{\mbox{\sc as}}(per(\mbox{\sc
  as}))$, where {\sc as} runs over those temporal rules without
preperiod. 
Apparently, after  $k\cdot \operatorname{PER}(v),\forall
k$, applications of  {\sc as}, we return to $v$, for all these rules
without preperiod simultaneously. 
After $k\cdot \operatorname{PER}(v)\pm 1$ steps, $\forall k$,
we are not at $v$ (unless the period is 1, and thus $v$ is a
fixed point). 
Hence, $v\mapsto v$ is impossible for all these timesteps
$k\cdot \operatorname{PER}(v)\pm 1$, and there is no such $\tau_0$. 

Finally, in the case that  $v$ is a fixed point under {\sc as}, no
cell changes its contents  
for this temporal rule and thus only $w=v$, but no $w\neq v$ is ever 
reached.\hfill $\Box$

{\it Definition~3.} \ \ 
We call an elementary cellular automaton pattern-universal at length
$n$, if it is able to convert any pattern $v$ in $\ff_2^n$ into any
other, {\it i.e.} satisfies property $(i)$ ($\forall v,w\in \ff_2^{n},
\exists \mbox{\sc as}\in{AS}_{n}, \exists \tau\in\nn: 
\mbox{\rm ECA}^\tau_{\mbox{\sc as}}(v) = w$).

If an ECA is pattern-universal for all $n\geq n_0$, for some $n_0$, it
is called uniformly pattern-universal.

{\it Corollary}\ \ 
ECA's from the $6$ families $19,23,37,41,57,$ and $105$
are pattern-universal for the lengths indicated in property $(i)$ above.

We conjecture that ECA's from families 19, 41, and 57 are uniformly
pattern-universal.

\section{The Finite Torus $\zz/n\zz$:
Functions}

In Section 2, we focussed on transductions $v\mapsto w$, which ---  in
general --- used different temporal rules for different $v$'s and
$w$'s, but for 
each pair $(v,w)$ stayed with the same rule, applied repeatedly. 

In this section, we are interested in functions 
$\ff_2^n\ni v\mapsto f(v)=w\in \ff_2^n$, which use the same temporal
rule sequence for all $v$, but ---  necessary to generate enough
variation --- concatenate several different temporal rules. 

We may restrict ourselves to ECA families 19, 23, 37, 41, 57, and 105.
Given a function on $\ff_2^n$ defined by the values $w(v)\in
\ff_2^n, \forall v$, our question is thus: 
$$\exists k\in\nn,
\exists\mbox{\sc as}_1,\dots \mbox{\sc as}_k\in{AS}_n, 
\forall v\in \ff_2^n:\mbox{\rm ECA}_{\mbox{\sc as}_k}(\dots(\mbox{\rm
  ECA}_{\mbox{\sc as}_1}(v))\dots) = w(v)?$$

\subsection{Bijective Functions}

We first consider bijective functions on $\ff_2^n$.
In this case the equivalent group-theoretic statement is:

{\it Do the $\mbox{\rm ECA}_{\mbox{\sc as}}\in{AS}_n$ 
$($written as permutations on the
set $\{0,1,\dots,2^n-1\})$ generate the full symmetric group
$S_{2^n}$} ? 

To answer this question, we used the program GAP (Graphs, Algorithms,
Programming) from RWTH Aachen (Prof.~Neub\"user's group) and
St.~Andrews University \cite{GAP}. Thank you! 

We ran GAP on some subsets of only 3 asynchronicity rules to show that 
 $\mbox{\rm ECA}_{\mbox{\sc as}}\in{AS}_n$ generates at least the
alternating group $A_{2^n}$, for $3\leq n\leq 11$.

Trying directly to obtain the group generated by the full set 
 $\mbox{\rm ECA}_{\mbox{\sc as}}\in{AS}_n$ overburdens GAP already from
$n=4$ on.
Therefore, in order to check for the generation of $S_{2^n}$, it is then
sufficient to exhibit at least one odd permutation, which is the case
for $n=3$, with the whole $S_{2^3}$ generated --- or to show that all
permutations generated by  $\mbox{\rm ECA}_{\mbox{\sc as}}\in{AS}_n$ are
even, which is the case for $4\leq n\leq 11$, and thus only
$A_{2^n}$, but not
$S_{2^n}$, is generated in these cases.

Out of the 6 ECA families satisfying property $(i)$, 
ECA-57 and ECA-105 are  the only ones, which have a locally bijective
update rule. 
Therefore, only these families must be considered. We immediately
have that temporal rules avoiding the symbol ``$\equiv$'' are bijective,
when the temporal rule is bijective, since different applications of
that temporal rule do not interfere with each other.  
On the other hand, for $n\neq 3$, all temporal rules involving the
symbol ``$\equiv$'' lead to non-bijective functions, see next subsection. 

The rules excluding $\equiv$ define bijective functions, whenever the
ECA itself is (locally) bijective, that is the application of such an
temporal rule for a single cell yields bijectivity.
Those temporal rules including $\equiv$ define the non-bijective functions.
Hence, the only way to generate bijective functions for $n\geq 4$ 
is by using ECA-57 or ECA-105,
and only applying temporal rules from $\{<,>\}^n$.

ECA-57: GAP tells us that the $2^n-2$ temporal rules from
$\{<,>\}^n\backslash\{<^n,>^n\}$ always yield at least the alternating
group $A_{2^n}$, which is in fact generated already by 3 of the
temporal rules, for $3\leq n \leq 10$.

The case $S_{2^n}$ vs.~$A_{2^n}$ is easiest checked by hand: 
Is there some odd permutation within the temporal rules? 
This is only the case for $n=3$.
For $4\leq n\leq 10$, all temporal rules yield even
permutations and thus can not generate the full $S_{2^n}$. 

Hence, for $n=3$, all bijective functions are generated through ECA-57
by concatenation of suitable temporal rules, while for $n\geq 4$, only
the even permutations from $A_{2^n}$ (that is half of the $2^n!$
bijective functions) are generated. 

ECA-105: GAP tells us that all bijective temporal rules combined generate
only fairly small groups: 
$<AS_3>$ has order 24,
$|<AS_4>| =   48 = 4!\cdot 2^1$,
$|<AS_5>| = 1920 = 5!\cdot 2^4$,
$|<AS_6>| = 11520 = 6!\cdot 2^6$,
$|<AS_7>| = 322560 = 7!\cdot 2^6$, all are far below $|S_{2^n}|=
2^n!$,
the number of bijective functions.

\subsection{Non-Bijective Functions}

We now turn to non-bijective functions.
Then $Im(f)\subset\ff_2^n$ with $|Im(f)|$ strictly less than $2^n$. 

We start with $n=3$.
The convex hull over all $\mbox{\sc{as}}\in{AS}_3$ has cardinality at
least $2^3!/2= 20160$ for the following ECA's, Table~3 
(the other ECA with bijective update rule,  ECA-105, generates only
344 functions): 

\begin{table}[h]
\centering
\bt{lr|lr}
ECA-25:&     22496&     ECA-46:&     89110  \\
ECA-110:&     23166&    ECA-41:&    210493  \\  
ECA-30:&     25258&     ECA-38:&    223102  \\
ECA-3:&     39155  &    ECA-27:&    268034  \\
{\bf ECA-57:}&{\bf 40320}&    ECA-35:&    751760  \\
ECA-11:&      52934  &  ECA-54:&   1.190.449\\
ECA-62:&     62683&     ECA-19:&   3.519.992\\
\et
\caption{Image size for ECAs on $\ff_2^3$}
\end{table}

There are $8^8$, about 16 Mio., functions on $\ff_2^3$. Hence, for
$n=3$, none of the ECA's even generates a quarter of all functions. 
The case ECA-57 is special in that this ECA actually generates all
bijective functions, but no non-bijective one, for $n=3$.

In the sequel, $n\geq 4$, we consider only ECA-57, which has sufficiently
many  bijective functions, namely $2^n!/2$, at least for $4\leq n
\leq 10$. We will generate a considerable subset of all functions by
suitably interleaving bijective and non-bijective temporal rules for
ECA-57.

We now consider ECA-57 for a temporal rule with a single $\equiv$ on 
$\zz/n\zz,n\geq 4$.

Considering larger neighborhoods, with 2 cells changing
simultaneously, also ECA-57 becomes non-surjective (we show the effect
of {\sc as}$_1=$  ``$\equiv$''  on the two middle cells for all
configurations of 4 adjacent cells): 

\begin{table}[h]
\centering
\bt{ccc}
$v$&$\mapsto$& ECA-57$(v)$\\
\cline{1-3}
0000&$\mapsto$&0110\\
0001&$\mapsto$&{\bf 0101}\\
1000&$\mapsto$&1110\\
1001&$\mapsto$&1101\\
0010&$\mapsto$&0000\\
0011&$\mapsto$&{\bf 0011}\\
1010&$\mapsto$&1100\\
1011&$\mapsto$&1111\\
0100&$\mapsto$&0010\\
0101&$\mapsto$&{\bf 0011}\\
1100&$\mapsto$&1010\\
1101&$\mapsto$&1011\\
0110&$\mapsto$&0100\\
0111&$\mapsto$&{\bf 0101}\\
1110&$\mapsto$&1000\\
1111&$\mapsto$&1001\\
\et
\caption{ECA-57: Effect of $\mbox{\sc as}_1=$ ``$\equiv$''}
\end{table}

We obtain the patterns 0011 and 0101 twice, while missing 0001 and
0111. Hence the image is smaller than the full $2^4$ by 2, or by a
factor of 7/8.

Extending this neighborhood of $\equiv$ to any size $n$, and using
only $<$ and $>$ for the other $n-1$ positions, before and after the
$\equiv$ transition, ECA-57 behaves bijectively.
Therefore, the whole image shrinks by just the factor 7/8, when
applying $\equiv$ once.

Since all temporal rules without $\equiv$ are bijective, and inclusion of
more than one $\equiv$ shrinks the image even further, we have the
following result on the functions that can be represented by ECA-57:

{\bf Theorem~3}

{\it
Let the patterns from $\{<,>\}^n\backslash \{<^n,>^n\}$ generate at
least the alternating 
group $A_{2^n}$ $($which is the case at least for $3\leq n\leq 10)$.

Let $f:\ff_2^n\to \ff_2^n, n\geq 4$ be any non-bijective function on at
least $4$ symbols.
Let $\#(w)=|\{v|f(v)=w\}|$ be the number of configurations $v$ leading
to configuration~$w$.
Then $f$ is representable by {\rm ECA-57} under asynchronicity, if and only if
$$\sum_{w\in \ff_2^n}\lfloor\#(w)/2\rfloor \geq 2^{n-3}.$$
}

{\it Proof.}
We first introduce the functions $\#$ on $\ff_2^n$ and $@$ on
$\nn_0$:

The multiplicity $\#(w)$ tells us, how often $w$ is reached, {\it
  i.e.} is the size of the preimage of $\{w\}$.

For $k\in\nn_0$, let $@(k)\in\nn_0$ be the number of results $w$
appearing with multiplicity $k$, $@(k) =|\{w\in \ff_2^n\colon
\#(w)=k\}|$.
In particular, $@(0)=2^n-|Im(f)|$ is the number of words avoided
by the image of $f$.
We have $\sum_kk\cdot @(k)=2^n$.

We make use of the  temporal rule $\mbox{\sc as}^*$ := ``$<\equiv
>\cdots>$'' which maps 
$2^{n-3}$ pairs $(v_1,v_2)$ onto  $2^{n-3}$ words $w$, and
otherwise is 1-to-1.
Hence, for $\mbox{\sc as}^*$, we have $@(1)=6\cdot 2^{n-3},@(0)=@(2)=2^{n-3}$.
We  generate $f$ by a chain
$f = \pi_k\circ \mbox{\sc as}^*\circ \dots\circ \pi_2\circ \mbox{\sc
  as}^*\circ \pi_1\circ 
\mbox{\sc as}^*$, alternating $\mbox{\sc as}^*$ and
permutations $\pi_k\in S_{2^n}$.

For the second and every further application of $\mbox{\sc as}^*$, we will
join $2^{n-3}-1$ words $v_1$ with $\#(v_1)>0$ to 
$2^{n-3}-1$ words $v_2$ with $\#(v_2)=0$, hence without changing 
the distribution $@$. 
We also map $6\cdot 2^{n-3}$ words 1-to-1, and
finally we join two multiplicities $\#(v_1),\#(v_2)$ by mapping
$v_1,v_2$ onto the same $w$,
the actual effect of this application of   $\mbox{\sc as}^*$.
The new values $@^+$ are thus 
$@^+(\#(v_1))=@(\#(v_1))-1,
@^+(\#(v_2))=@(\#(v_2))-1,
@^+(\#(v_1)+\#(v_2))=@(\#(v_1)+\#(v_2))+1,$ and 
$@^+(k)=@(k)$ otherwise. In this way, we eventually arrive at a
distribution $@$ as required by $f$.

To achieve this, we permute values in between applications of
$\mbox{\sc as}^*$. 
In this (slow) way, we eventually get to the distribution of $\#(w)$
required by $f$.

The final permutation $\pi_k$ maps the $v$ with multiplicities $\#(v)>0$ 
to the correct values $w\in Im(f)$.

Since we always have two words $v_1,v_2$ mapping to the same $w$ under 
$\mbox{\sc as}^*$, and also two words $w_1,w_2$ outside $Im(\mbox{\sc as}^*)$,
any $\pi_k\in S_{2^n}\backslash A_{2^n}$ can be extended by one of the
transpositions $(v_1,v_2)$ or $(w_1,w_2)$ to an equivalent $\pi_k'\in
A_{2^n}$.  

Concerning the ``only if'' part, already the first application of 
$\mbox{\sc as}^*$ would decrease the number of values below $|Im(f)|$.
\hfill $\Box$

\subsection{Examples}

The superscript ${}^{(n)}$ indicates the torus size.

\ul{INC}$^{(3)}$\ \ For $n=3$, let $w= v+1\mod
8$. This is an odd bijective 
function, and hence representable for this $n=3$.

\ul{MUL-BY-3}$^{(3)}$\ \ For $n=3$, let $w= 3\cdot v\mod 8$. Same as with INC.

\ul{MUL-BY-2}$^{(3)}$\ \ For $n=3$, let $w= 2\cdot v\mod 8$. From
$2\cdot 0 = 2\cdot 4 = 0\mod 8$, this function is not bijective, and
hence not representable by ECA-57 for $n=3$. 

\ul{INC}$^{(4)}$\ \ For $n=4$, let $w= v+1\mod 16$. 
As with $n=3$, this is an odd bijective function.
Contrary to the case $n=3$, a representation by ECA-57 is not possible
for  $n\geq 4$.

\ul{INC'}$^{(4)}$\ \ For $n=4$, let $w= v+1\mod 15, 15\mapsto 15$.
This is an even bijective function, and thus representable.

\ul{MUL-2-BY-2}$^{(4)}$\ \ For $n=4$, let $v= a|b$, $0\leq a,b\leq 3$ and $w =
a\cdot b$.
The range is given by the multiset $\{0^7,1,2^2,3^2,4,6^2,9\}$, where
superscripts show the number of occurrences.
The sum $\sum_{w\in\ff_2^4}\lfloor\#(w)/2\rfloor= 6\geq 2^{n-3} = 2$
is large enough (the range is sufficiently small thus) to allow
shrinking by {\it e.g.} repeated application of the asynchronicity
pattern ``$<=>>$'' and suitable permutations. 
Multiplication can thus be computed by ECA-57 through asynchronicity. 
{\it  How} to do it exactly, is a more complicated case, see Open Problems.

\ul{MUL-k-BY-k}$^{(2k)}$: Zero appears $2\cdot 2^k-1$ times, and 
$1\leq a< b\leq 2-1$ yields $a\cdot b= b\cdot a$ that is at least
${2^k-1\choose 2}$ pairs.
Hence, we have
$\sum_{w\in\ff_2^4}\lfloor\#(w)/2\rfloor
\geq 2^k-1 + (2^k-1)\cdot (2^k-2)/2 > 2^{2k-3}= 2^{n-3}$ (with $k\geq
2$). 
All these multiplications can therefore be computed by ECA-57, using
asynchronicity. 

\newpage
Boolean and arithmetic functions on $k$ bits, $n = 2k$:

Let $v=a|b$ with $0\leq a,b< 2k$.
Then 
$\ul{f_{\lor,1}}(v) = (0|a\lor b)$,
$\ul{f_{\lor,2}}(v) = (a\lor b|a\lor b)$,
$\ul{f_{\land,1}}(v) = (0|a\land b)$,
$\ul{f_{\land,2}}(v) = (a\land b|a\land b)$,
$\ul{f_{\oplus,1}}(v) = (0|a\oplus b)$,
$\ul{f_{\oplus,2}}(v) = (a\oplus b|a\oplus b)$,
$\ul{f_{-}}(v) = (a - b\mod 2^{2k})$,
$\ul{f_{-'}}(v) = (0|(a- b)\mod 2^k)$,
can all be computed by ECA-57 under asynchronicity, for any $k$ that is
any even $n$: 

All these Boolean functions are commutative, $a\circ b = b\circ a$,
and thus enough pairs $(v_1,v_2)$ with $f(v_1)=f(v_2)$ exist
to have $\sum \lfloor \#(w)/2\rfloor \geq 2^{2k-3}$.

For $0\leq v< 2^n$, let
$\ul{f_{\operatorname{NEG}}}(v) = (-v)$ (2's complement). This is an odd
bijection with the two fixed points $0$ and $2^{n-1}$, and $2^{n-1}-1$
transpositions, hence  computable by ECA-57 only for  $n=3$. 

$\ul{f_{\operatorname{COMP}}}(v) = (v \oplus 111\cdots 111)$
(1's complement), on the other hand, is an even bijection, 
computable for all $n$.

{\bf Example~5}\ \ 
A detailed description of the calculation of $\operatorname{INC}^{(3)}$ and
$\operatorname{MUL-BY-3}^{(3)}$.
The left column indicates the temporal rule and partition of $\zz/3\zz$

\begin{table}[h]
\centering
$\ba{lcccccccc}
 \multicolumn{9}{c}{\operatorname{INC}^{(3)}}
\\
\cline{1-9}
&000&001&010&011&100&101&110&111
\\
 ^{<><(1|0|2)}&011&101&100&110&111&010&001&000
\\
 ^{<>>(1|2|0)}&010&110&011&000&001&101&100&111
\\
 ^{<>\equiv(1|0,2)} &101&000&110&011&100&010&111&001
\\
 ^{\equiv><(0,1|2)}&010&111&100&110&011&001&000&101
\\
 ^{>><(0|1|2)}&001&010&011&100&101&110&111&000
\\
\ea$

\noindent
$\ba{lcccccccc}
\multicolumn{9}{c}{\operatorname{MUL-BY-3}^{(3)}}\\
\cline{1-9}
&000&001&010&011&100&101&110&111\\
^{<<>(2|1|0)}&101&010&111&100&110&011&001&000\\
^{><>(2|0|1)}&001&101&100&010&011&000&110&111\\
^{\equiv<>(2|0,1)}&110&011&010&111&000&101&001&100\\
^{>\equiv<(0|1,2)}&101&100&001&010&110&000&111&011\\
^{>><(0|1|2)}&000&011&110&001&100&111&010&101\\
\ea$
\caption{$\operatorname{INC}^{(3)}$ and
$\operatorname{MUL-BY-3}^{(3)}$ in detail}
\end{table}

\section*{Further Research and Open Problems}

1. Give an algorithm to calculate the temporal sequence $(\mbox{\sc
  as}_1,\mbox{\sc as}_2,\dots,\mbox{\sc as}_k)$ 
for a function on $\ff_2^n$ directly from the function values, given
{\it e.g.} as permutation on $\{0,1,\dots,2^n-1\}$, instead of
searching through the full tree ${AS}_n^k$.

2. Consider temporal sequences that do not depend on the position, but
on the rule to be applied, {\it e.g.} 
first update at all corresponding sites $000\mapsto p_0$ , then 
$110\mapsto p_6$, then
$010\mapsto p_2$ etc.
There are 8! = 40320 such temporal rules, independent of $n$.

How do we treat actions that already had their turn, but whose
neighborhood only turns up later? 
Update immediately upon creation, never in this round, ...?

This could mimic chemical reactions, {\it e.g.} in cell biology, DNA
expression, where some reactions are faster than others, depending on
their reaction rate constant $k$.

3. As in 2., but associate a latency time with each temporal rule: As
soon as the corresponding neighborhood pattern is created, wait for
its latency time and then update according to the temporal rule.

4. What can we say about the alphabet $\{0,1,2\}$ instead of
$\{0,1\}$?

There are now $3^{3^3} \approx 2^{43}$ ECA's to be considered. 
Since there are $3^n-2^{n+1}+2=\Theta(3^n)$ asynchronicities (Definition~1) and
exactly $3^n$ configurations on $\zz/n\zz$, 
an analogue of properties $(o)$ to $(iv)$ is now impossible due to
lack of temporal rules. 
However, we may ask, how the number of configurations actually reached
grows with $n$.  
Do we ever obtain the full diversity of $3^n-2^{n+1}+2$ results?

\section*{Conclusion}

We have introduced temporal order via temporal rules as a means to
diversify the behaviour of elementary cellular automata. 

In particular, ECA's with update rules 19, 41,   and 57  are pattern
universal for $n\leq 12$, 
achieving any desired pattern transduction $v\mapsto w$, applying
iteratedly a single temporal rule. 
We conjecture that they are indeed uniformly pattern-universal.

ECA-57 produces any even (as permutation)  bijective
function on $\{0,1\}^n$, for $n\leq 10$, and all non-bijective ones
that join at least $2^{n-3}$ pairs of argument values.

{\bf 
Temporal order is thus a third way to encode information and
algorithms, after programs (ECA's) and data (initial
configurations).} 

This may have farreaching consequences, {\it e.g.} for modeling gene
expression, since physico-biological processes seldomly achieve exact
synchronicity.

\newpage
\section*{Appendix A -- ECA Families}

{\small{
Each family (equivalence class under the symmetries 0/1 and L/R
($c_{i-1}\leftrightarrow c_{i+1}))$ 
consists in up to 4 ECAs with numbers 
ECA={\it abcdefgh}$_2$,\\
ECA$\stackrel{L/R}{\longleftrightarrow}$
 {\it aecgbfdh}$_2$,
ECA
$\stackrel{0/1}{\longleftrightarrow}$
$\overline{\mbox{\it hgfedcba}}_2$, 
and
ECA$\stackrel{L/R, 0/1}{\longleftrightarrow}
 \overline{\mbox{\it hdfbgcea}}_2$.

\noindent
\bt{llll}
0 (255),& 
1 (127),&
2 (191 16 247),&
3 (63 17 119),\\
4 (223),&
5 (95),&
6 (159 20 215),&
7 (31 21 87)\\
8 (239 64 253),&
9 (111 65 125),&
10 (175 80 245),&
11 (47 81 117),\\
12 (207 68 221),&
13 (79 69 93),&
14 (143 84 213),&
15 (85),\\
18 (183),&
19 (55),&
22 (151),&
23,\\
24 (231 66 189),&
25 (103 67 61),&
26 (167 82 181),&
27 (39 83 53),\\
28 (199 70 157),&
29 (71),&
30 (135 86 149),&
32 (251),\\
33 (123),&
34 (187 48 243),&
35 (59 49 115),&
36 (219),\\
37 (91),&
38 (155 52 211),&
40 (235 96 249),&
41 (107 97 121),\\
42 (171 112 241),&
43 (113),&
44 (203 100 217),&
45 (75 101 89),\\
46 (139 116 209),&
50 (179),&
51,&
54 (147),\\
56 (227 98 185),&
57 (99),&
58 (163 114 177),&
60 (195 102 153),\\
62 (131 118 145),&
72 (237),&
73 (109),&
74 (173 88 229),\\
76 (205 76 205),&
77,&
78 (141 92 197),&
90 (165 90 165),\\
94 (133),&
104 (233),&
105,&
106 (169 120 225),\\
108 (201),&
110 (137 124 193),&
122 (161),&
126 (129),\\
128 (254),&
130 (190 144 246),&
132 (222),&
134 (158 148 214),\\
136 (238 192 252),&
138 (174 208 244),&
140 (206 196 220),&
142 (212),\\
146 (182),&
150,&
152 (230 194 188),&
154 (166 210 180),\\
156 (198),&
160 (250),&
162 (186 176 242),&
164 (218),\\
168 (234 224 248),&
170 (240),&
172 (202 228 216),&
178,\\
184 (226),&
200 (236),&
204,&
232
\et

}}
\end{document}